\documentclass[aps,pra,twocolumn,amsfonts,amssymb,amsmath,showpacs,
floatfix,nofootinbib,groupedaddress,superscriptaddress,citesort]{revtex4}
\usepackage{mathrsfs}
\usepackage{amsfonts}
\usepackage{amstext}
\usepackage{amsmath}
\usepackage{amssymb}
\usepackage{bm}% bold math
\usepackage{bbm}
\usepackage[dvips]{graphicx}
\def\qed{\leavevmode\unskip\penalty9999 \hbox{}\nobreak\hfill
     \quad\hbox{\leavevmode  \hbox to.77778em{%
              \hfil\vrule   \vbox to.675em%
               {\hrule width.6em\vfil\hrule}\vrule\hfil}}
     \par\vskip3pt}

\begin{document}
\title
{Measure-independent description of wave-particle duality  via coherence}

\author{Zhaofang Bai}
\email{baizhaofang@xmu.edu.cn} \affiliation{School of Mathematical
Sciences, Xiamen University, Xiamen, Fujian, 361000, China}

\author{Shuanping Du}\thanks{Corresponding author}
\email{dushuanping@xmu.edu.cn} \affiliation{School of Mathematical
Sciences, Xiamen University, Xiamen, Fujian, 361000, China}

%\thanks{{\it PACS numbers: 03.67.HK, 02.10.Yn, 03.65.Aa}}
%\thanks{{\it Mathematics Subject Classification: 47B49, 46L07, 47L07, 46N50}}
%\thanks{{\it Key words and phrases.} strictly incoherent operations, frozen coherence, the $l_1$-norm of coherence
%}

%\thanks{This paper is in final form and no version of it will be
%submitted for publication elsewhere.}

\begin{abstract}
Wave-particle duality as one of the expression of Bohr complementarity is a significant concept in the field of quantum mechanics.
Quantitative analysis  of wave-particle duality aims to establish a complementary relation between the particle and wave properties.
Beyond the conventional quantitative analysis depending on special choice of quantum information measures,
we are aimed to provide a  measure-independent complementary relation  via coherence. By employing maximally coherent states
in the set of all states with fixed diagonal elements, a measure-independent complementary relation is proposed. Based on this, we  give a measure-independent description of wave-particle-mixedness triality in $d$-path interferometers. Our complementary relations  reveal the relationship between wave-particle duality and quantum coherence, and also give a justification to coherence as it truly brings out the wave nature of quantum systems at its heart.

\end{abstract}


\maketitle

${Introduction}.$  Wave-particle duality (WPD) is a striking physical phenomenon
to exhibit the curious nature of quantum mechanics and has taken a profound influence on the development of quantum mechanics \cite{Bohr,Mandel,Ficek}.
The particle properties are characterized by how much
information one has about which path the particle took
through the device. The wave properties determine the
visibility of the interference pattern. There is a complementary relationship
between the particle and wave properties: the stronger one
is, the weaker the other is.

A quantitative version of the complementary relationship for interferometers with two internal paths is initially
investigated by Wootters and Zurek \cite{Wootters}.  The relationship was put into an
elegant form by Greenberger and Yasin \cite{Greenberger},  $$D^2+V^2\leq 1,\ \eqno{(1)}$$
 $D$ is a measure of path information and $V$ is the
visibility of the interference pattern.

This is an experimentally testable inequality, as it involves
physically measurable quantities.
The work was pushed ahead by Jaeger et al. \cite{Jaeger}, who provide
the complementary relationship for interferometers with more than two paths.
Englert introduced detectors into the problem and derived a duality relation between this type of path
information and the visibility that took the form of Eq. (1) \cite{Englert1}.

With the flourishment  of quantum resource theory,    there  are  grown interests in  revealing the relationship between WPD and various quantum concepts, such as entanglement \cite{Jakob1,Jakob2,Mafei}, entropic uncertainty \cite{Coles1,Coles2}, quantum state discrimination \cite{Lu}, and coherence \cite{Durr,Bera,Bagan,Luo,Bu,Roy,luo2,Kim,Fei}. However, most of which depend on some special choice of quantum information measures out of computing needs,  such as the entropy \cite{Coles1}, the $l_1$-norm and the  relative entropy of coherence \cite{Bagan}, the Wigner skew information \cite{luo2}.                                                                                                                                                                                                         Our finding is  that the description of WPD actually can be made
independent of any special choice of coherence measurers \cite{Baumgratz} and  a unified description of WPD via coherence is provided in this paper.
This is harmonic with the fact that WPD  is a measure-independent physical phenomenon. Based on this result, we  propose a measure-independent description of wave-particle-mixedness triality in d-path interferometers.
These results not only allow us to formulate a simple criterion
to quantifying quantum wave and particle feature, but also show that the axiomatic definition of coherence measures \cite{Baumgratz} is realistic from the viewpoint of quantifying WPD.
\vspace{0.1in}

\begin{figure}[htbp]
	\centering
\includegraphics[scale=0.12]{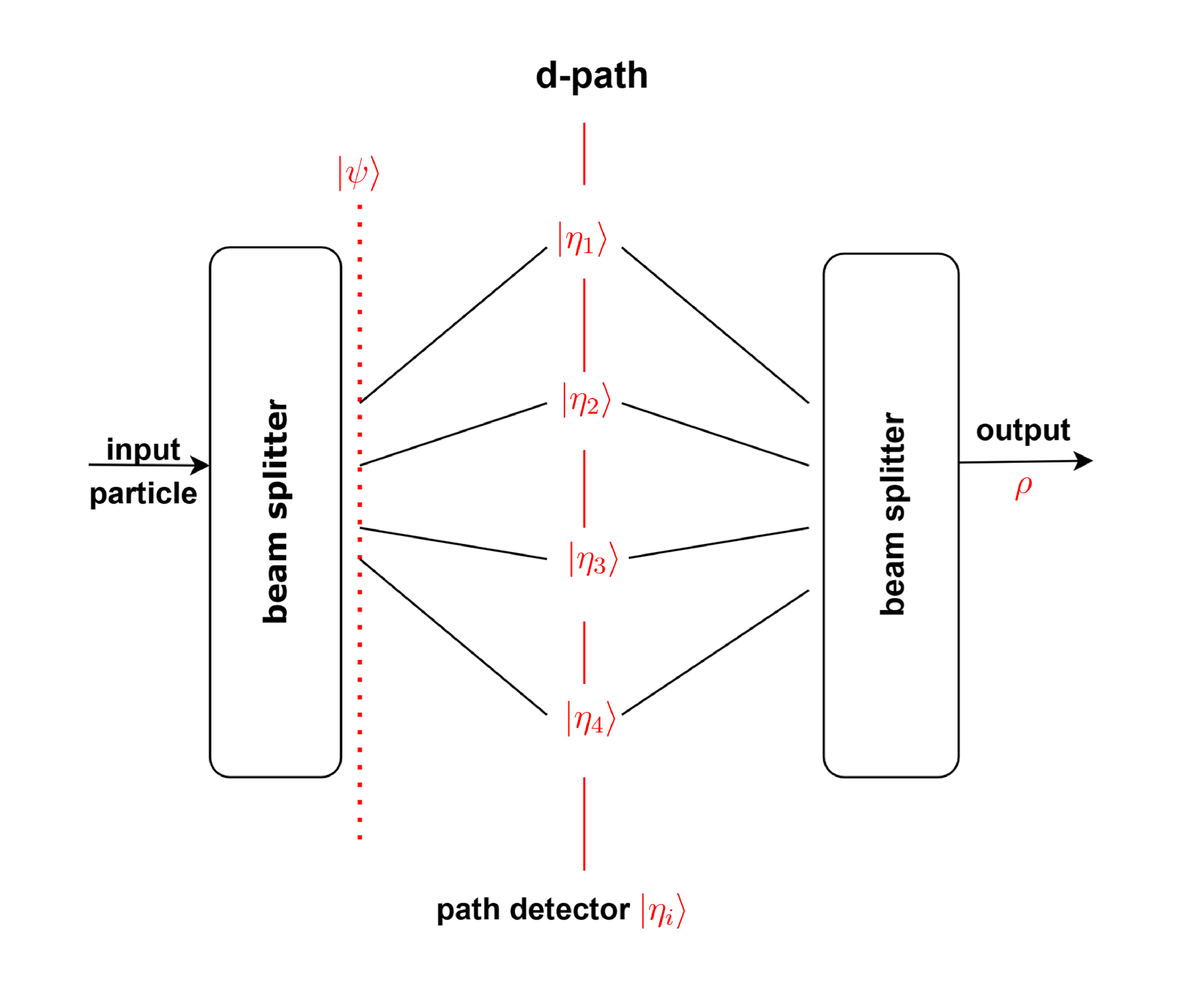}
 \caption{\small Diagram of a $d$-path interferometer with the path detector. }
\end{figure}

$Wave-particle\  duality.$ Let us recall the mechanism of $d$-path interferometers with path detectors (see Fig. 1).
 We start with pure quantons entering a  $d$-port interferometer via a generalized beam splitter which makes it in the superposition state:
$$|\psi\rangle=\sum_{i=1}^d \sqrt{p_i}|i\rangle.\eqno {(2)}$$ The orthogonal basis $\{|i\rangle\}_{i=1}^d$ denotes the $d$ possible paths, and $\sqrt p_i$ is the amplitude taking the $i$-th path.
While in the interferometer, the particle interacts with another system, called the detector. The detector starts in a global state $|\eta_0\rangle$ and changes from $|\eta_0\rangle$ to $|\eta_i\rangle$ $(1\leq i\leq d)$ when the particle passes through the $i$-th path. After the particle has interacted with the detector $|\eta_0\rangle$,  the state of entire system is $$|\Psi\rangle=\sum_{i=1}^d \sqrt{p_i}|i\rangle|\eta_i\rangle.\eqno {(3)}$$ By tracing out the detector in $|\Psi\rangle$, the density matrix of the particle is given by
$$\rho=Tr_{det}(|\Psi\rangle\langle\Psi|)=\sum_{1\leq i,j\leq d} \sqrt{p_ip_j}\langle\eta_i|\eta_j\rangle|i\rangle\langle j|,\eqno {(4)}$$ this will be the state hits the screen.

Following the requirements for the quantification of quantum wave and particle feature \cite{Durr,Peng,Englertx}, the quantification of the quantum wave feature (denoted  by $W(\rho)$) should satisfying the following criteria:

(i) $W(\rho)=0$ if the  state $\rho$ is diagonal in the prefixed basis $\{|i\rangle\}_{i=1}^d$.

(ii) $W(\rho)=1$  if $\rho$ is  pure and   $\langle i| \rho|i\rangle=\frac{1}{ d}$ for every  $1\leq i\leq d$.

(iii)  $W(\rho)$ is invariant under permutations of the $d$-path labels.

(iv)   $W(\cdot)$  is convex.

Correspondingly, the particle aspects  (denoted  by $P(\rho)$) should have the following criterias \cite{Durr,Englertx}:

(i) $P(\rho)=1$   if $\langle  i|\rho|i\rangle=1$  for some $i$.

(ii) $P(\rho)=0$  if $\langle  i|\rho|i\rangle=\frac{1}{ d} $ for every  $1\leq i\leq d$.

(iii)  $P(\rho)$ is  invariant under permutations of the $d$-path labels.

(iv)   $P(\cdot)$  is convex.

We remark that all the above  mathematical requirements are motivated by intuitive theoretical and experimental considerations \cite{Durr,Peng,Englertx}.
\vspace{0.1in}

$Coherence \ resource \ theory.$
To present our finding clearly, let us recall the
standard formalism of the coherence resource theory \cite{Baumgratz}. There are two fundamental elements of the coherence resource theory,
namely, incoherent states and incoherent operations.  All diagonal states in a fixed basis $\{|i\rangle\}_{i=1}^d$
are defined as incoherent states,  that is, $$\rho=\sum_{i=1}^d\lambda_i|i\rangle\langle i|, \eqno{(5)}$$ $\lambda_i\geq 0$ and $\sum_{i=1}^d{\lambda_i}=1$.
 The family of incoherent states is denoted by ${\mathcal I}$. All non-diagonal states  in
the basis are  named coherent states.
Every incoherent operation  is
specified by a set of operation elements $\{K_l\}$ such that $K_l\rho K_l^\dag/Tr(K_l\rho K_l^\dag)\in {\mathcal
I}$ for all $\rho\in {\mathcal I}$, $$\Phi(\rho)=\sum_{l}K_l \rho K_l^\dag.\eqno{(6)}$$  Such operation elements $\{K_l\}$ are called incoherent.
This definition of incoherent operations ensures that in an overall quantum operation $\rho\mapsto \sum_lK_l\rho K_l^\dag$, even if one does not have access to
individual outcomes $l$, no observer would conclude that coherence has been generated from an incoherent state.

Based on Baumgratz et al's suggestions \cite{Baumgratz},
any proper coherence measure $C$ is a non-negative function on quantum states and
must satisfy the following conditions:

$(C1)$ $C(\rho)\geq 0$ for each density operator $\rho$, and $C(\rho)= 0$ for $\rho\in{\mathcal I}$;

$(C2a)$ $C(\rho)$ is decreasing under every incoherent operation $\Phi$:
$C(\Phi(\rho))\leq C(\rho)$,

or $(C2b)$ $C(\rho)$ is decreasing under selective measurements on average: $\sum_l p_l C(\rho_l)\leq C(\rho)$
for all $\{K_l\}$, where $\{K_l\}$ is from the operation representation of $\Phi$, $\rho_l=\frac{K_l\rho
K_l^\dag}{p_l}$ and $p_l={\rm Tr}(K_l\rho K_l^\dag)$.

$(C3)$ $C(\rho)$ is convex:
$C(\sum_l p_l\rho_l)\leq \sum_l p_l C(\rho_l)$ for any set of states
$\{\rho_l\}$ and any $p_l\geq 0$ with $\sum_l p_l=1$.\\

Based on above mentioned postulates, there are two important coherence measures named the $l_1$-norm
of coherence and the  relative entropy of coherence have been introduced \cite{Baumgratz}.
The $l_1$-norm
of coherence  is defined as the sum of the absolute values of all
the off-diagonal elements $\rho_{ij}$ of $\rho$ , i.e., $$C_{l_1}(\rho)=\sum_{1\leq i\neq i\leq d}|\rho_{ij}|. \eqno {(7)}$$
Using this measure, we define the normalized $l_1$-norm of coherence as ${\mathcal C}_{l_1}(\rho)=\frac{C_{l_1}(\rho)}{ d-1}$. The  relative entropy of coherence is defined as $$C_{Re}(\rho)=\min_{\sigma\in{\mathcal I} } S(\rho||\sigma),\eqno {(8)}$$ where $S(\rho||\sigma)=Tr(\rho\log\rho-\rho \log\sigma)$ is quantum relative entropy \cite{Baumgratz}. We also consider the normalized relative entropy of coherence by ${\mathcal C}_{Re}(\rho)=\frac{C_{Re}(\rho)}{\log d}$, here $\log d$ denotes logarithms to base $2$. Besides the $l_1$-norm
of coherence and the  relative entropy of coherence, various important  coherence measures  are introduced, such as
convex roof construction of coherence \cite{Du}, the geometric coherence \cite{Streltsov}, the coherence cost \cite{Yuan,Winter}, and the robustness of coherence \cite{Napoli,Piani}.

\vspace{0.1in}

$Coherence \ and \ wave \ features.$
Quantum interference, which has its origin in the superposition principle, is a basic ingredient of WPD. Traditionally, it is often characterized via the fringe
visibility from the an operational perspective. Coherence is a fundamental feature of quantum physics, which signifies the possible superposition between orthogonal quantum states. Again, it is largely believed that quantum coherence has a strong correspondence with the wave nature of a quantum particle.
In particular, with the flourishing of quantum information theory, there are growing interests in exploring the role of interference in information processing and their relations with  coherence.
For instance, the ability of an operation to create coherence is related to the performance of interference experiment in ref. \cite{Masini}. By the $l_2$-norm of coherence, the duality relations of WPD in $n-$path $(n\geq 2)$ interferometers has been proposed by Durr \cite{Durr}.
Expressing the density matrix of the particle inside the interferometer in a path basis, his measure of path information depended on the diagonal elements of the density matrix and his measure of visibility depended on the off-diagonal elements.  The relationship between the $l_1$-norm of coherence and WPD for arbitrary multipath quantum interference phenomena has been revealed in \cite{Bera}. Bagan et al. have derived two dual relations relating the path information about a particle inside a multipath interferometer to the $l_1$-norm of coherence and the relative entropy of coherence \cite{Bagan}. Applying quantifiers of coherence
by the convex roof construction, the authors have built the complementary relationship between WPD and mixedness in $n$-path interferometers \cite{Kim}.
These duality relations then give a justification to the measure of quantum coherence as it truly brings out the wave nature of quantum particle.
However, these duality relations depend on some special choice of coherence measures out of computing needs, an interesting question is how to derive a
measure-independent description of WPD via coherence \cite{Kim}.

\vspace{0.1in}

$Measures  \ of \ different \ quantum \ features.$
A major challenge in building a measure-independent complementary relationship
is how to meet computing needs for all coherence measures. %We find maximally coherent
%states in the set of all states with fixed diagonal elements is an effective instrument to overcome the challenge.
In the following, we state our motivation for measures of different  quantum features.
With the development of coherence resource theory, the wave property is characterized by off-diagonal elements of a density matrix $\rho$, and the particle property is determined by diagonal elements of $\rho$ \cite{Durr, Peng, Englertx, Angelo, Fu2}.
Motivated by the observation, the wave feature of $\rho$ can be measured by $C(\rho)$ and the particle feature $D(\rho)$ of the mixed state $\rho$ ought to be described by
$D(|\rho\rangle\langle\rho|)$, here the pure state $$|\rho\rangle=\sum_{i=1}^d\sqrt{\rho_{ii}}|i\rangle \ \eqno{(9)}$$
has the same main diagonal elements as $\rho$. Because the WPD relations have neatly complementary relations in the pure states,
out of intuition, we define
$$D(\rho)=D(|\rho\rangle\langle\rho|)=1-C(|\rho\rangle\langle\rho|), \ \eqno{(10)}$$ as a measure of the  particle property of a general mixed state $\rho$. A direct computation shows that  $|\rho\rangle$ bears the following properties:

(i) $|U\rho U^\dag\rangle\langle U\rho U^\dag|=|\rho\rangle
 \langle\rho|$ for any diagonal unitary matrix $U$.

(ii) $C(|U\rho U^\dag\rangle\langle U\rho U^\dag|)=C(|\rho\rangle
 \langle\rho|)$ for any permutation unitary matrix.

(iii) If $\rho$ is pure, then $|\rho\rangle\langle\rho|=U\rho U^\dag $ for some diagonal unitary matrix $U$.

\vspace{0.1in}
${Results}.$ Now, we give the main  results of this note.

\vspace{0.1in}
\textbf{Theorem 1.} {\it For any  normalized coherence measure $C$ and mixed state $\rho$, $$C(\rho)+D(\rho)\leq 1, \ \eqno{(11)}$$
where $C(\rho)$ quantifies the quantum wave feature,
$D(\rho)=1-C(|\rho\rangle\langle\rho|)$ quantifies the quantum particle
feature, the equality holds if $\rho$ is pure.}
\vspace{0.1in}

{\bf Proof.} The key step is to show  $C(|\rho\rangle\langle\rho|)-C(\rho)\geq 0$.
For a fixed probability distribution $\{\rho_{ii}\}_{i=1}^d$,  let $$S=\{\sigma=(\sigma_{ij}):\sigma_{ii}=\rho_{ii}, i=1,2, \ldots, d\},\eqno(12)$$
a maximally coherent state of $S$  means a state from which all other states of $S$ can be created via incoherent operations.
Based on the majorization condition of determining convertibility from pure states to mixed states \cite[Theorem 2]{Du2}, we can obtain that the pure state $|\rho\rangle=\sum_{i=1}^d \sqrt {\rho_{ii}}|i\rangle$ is maximally incoherent in $S$,
 i.e., there exists an incoherent operation $\Phi$ such that $\Phi(|\rho\rangle\langle\rho|)=\rho$. Monotonicity of coherence measures under incoherent operations shows $C(|\rho\rangle\langle\rho|)\geq C(\rho)$, so $C(|\rho\rangle\langle\rho|)-C(\rho)\geq 0$. Next, we will show that $C(\cdot)$ satisfies all conditions of wave feature.
By the property (C1) of coherence measures, $C(\rho)=0$ for any diagonal state $\rho$. Note that $\rho$ is maximally incoherent under incoherent operations if $\rho$ is pure and a uniform superposition of the states in the computational basis \cite{Baumgratz},  thus $C(\rho)$ reaches its global maximum.  For every permutation matrix $U$, it is evident $\rho\mapsto U\rho U^\dag$ is an incoherent operation. From the
the monotonicity of coherence measures under incoherent operations, we know that $C(\rho)$ is invariant under permutations of diagonal elements of $\rho$. The convexity of $C(\cdot)$ is from (C3).
Hence $C(\cdot)$ possesses  all properties for quantifying  wave feature.
Note that  the restriction of any coherence measure (satisfying C1,C2b and C3) to
pure states can be derived by some symmetric concave function \cite{Du},  a direct check shows $D(\cdot)$ satisfies all properties for quantifying particle feature.\ $\square$
\vspace{0.1in}

By Theorem 1, one can see that both the wave feature and particle feature of a quantum system can be quantified via coherence measures.
We remark that the wave and particle nature of a quantum system was observed
long back, while quantification of the wave and particle nature  via coherence measure was presented only a few years back \cite{Bera,Bagan,Luo,Bu,Roy,luo2,Kim,Fei}. The  complementary relations given in literatures have specific forms based on special coherence measures. A natural question is that whether there exists a generic complementary relation which is suitable for all coherence measures \cite{Kim}.
Theorem 1 provides a desired complementary relation which can be applied to all coherence measures. Diagrams of complementary relations between wave and particle properties will be contained in Fig. 1 and Fig. 2. Let $D=\sqrt{D(\rho)}, V=\sqrt{C(\rho)}$, then the relation (11) is exactly of the form given in (1).

Choosing $C$ as a special coherence measure from the convex roof construction \cite{Du}, the  relation (11) becomes  the neat duality relation proposed recently by Y. Tsui and S. Kim \cite{Kim}: $$C_f(\rho)+D_f(\rho)\leq 1,\eqno {(13)}$$ where $C_f(\rho)=\min_{p_j,|\psi_j\rangle}\sum_jp_jf(|\psi_j\rangle)$ is the quantification of
the quantum wave feature, $D_f(\rho)=1-C_f(|\rho\rangle\langle\rho|)$ is the quantification of the quantum particle
feature, $f$ is some normalized symmetric concave function. In addition, the equality holds if $\rho$ is pure.

Choosing ${\mathcal C}$ as  the normalized $l_1$-norm of  coherence and the normalized relative entropy of coherence respectively, the relation (11) can induce                                                 complementary relations: $${\mathcal C}_{l_1}(\rho)+ 1-{\mathcal C}_{l_1}(|\rho\rangle\langle\rho|)\leq 1, \eqno {(14)}$$
$${\mathcal C}_{Re}(\rho)+ 1-{\mathcal C}_{Re}(|\rho\rangle\langle\rho|)\leq 1, \eqno {(15)}$$
here $${\mathcal C}_{l_1}(\rho)=\frac{1}{d-1}\sum_{1\leq i\neq j\leq d}\sqrt{p_i p_j}|\langle\eta_j|\eta_i\rangle|,$$ $${\mathcal C}_{l_1}(|\rho\rangle\langle\rho|)=\frac{1}{d-1}\sum_{1\leq i\neq j\leq d}\sqrt{p_i p_j},$$ $${\mathcal C}_{Re}(\rho)=-\log d(\sum_{i=1}^d p_i\log{p_i}+S(\rho)),$$  $${\mathcal C}_{Re}(|\rho\rangle\langle\rho|)=-\log d\sum_{i=1}^d p_i\log {p_i},$$ $S(\rho)$ is the von Neumann entropy. In fact, two relations of WPD under the $l_1$-norm of coherence and the relative entropy of coherence have been given by E. Bagan et al. \cite{Bagan}:
$$({\mathcal C}_{l_1}(\rho))^2+(\frac{dP_s-1}{d-1})^2\leq 1, \eqno {(16)}$$
$$\frac{C_{Re}(\rho)}{H(\{p_i\})}+\frac{Acc(D)}{H(\{p_i\})}\leq 1, \eqno {(17)}$$ where $$P_s=\max_M\sum_{i=1}^d p_i tr(\Pi_i|\eta_i\rangle\langle\eta_i|),$$
$$Acc(D)=\max_M H(M,D),$$ $H(\{p_i\})=-\sum_{i=1}^d p_i\log_2^{p_i}$ is the Shannon entropy, $M$ is a positive operator value measure with elements $\Pi_i\geq0$, which satisfy $\sum_{i=1}^d\Pi_i=I$, $$H(M,D)=H(\{p_i\})+H(\{q_i\})-H(\{p_{ij}\}),$$ $$q_i=tr(\Pi_i\sum_{i=1}^dp_i|\eta_i\rangle\langle\eta_i|),$$ $$p_{ij}=tr(\Pi_i|\eta_j\rangle\langle\eta_j|)p_j.$$ One key point of (16) (respectively (17)) is to find  the optimal solution of $P_s$ $(Acc(D))$. The solution to $P_s$ is known in complete generality for two general states \cite{Hels}, but only in special case for more than two states, while the solution to  $Acc(D)$ is totally unknown.
Therefore, compared with the two known complementary relations (16) and (17) from ref. \cite{Bagan}, our complementary relations (14) and (15) are  easily accessible for two reasons: (i) there is no optimal process involved, (ii) it is independent of positive operator value measure.
A pictorial representations of  (14) and (15),  i.e., complementary relations between wave and particle properties under the $l_1$-norm and the relative entropy of coherence,  will be shown in  Fig. 2 and Fig. 3 respectively.

\vspace{0.1in}

From Theorem 1, we  can infer a nice complementarity relation of wave-particle-mixedness triality which is also independent of any special choice of coherence measurers. In order to obtain a neatly complementary relation of WPD, we set $$M(\rho)=C(|\rho\rangle\langle\rho|)-C(\rho). \ \eqno{(18)}$$
Combining properties (ii) and (iii) of $|\rho\rangle$, we have $M(\rho)=0$ for pure state $\rho$. Thus $M(\cdot)$ may
quantifies the mixing of a quantum state $\rho$. In addition, $M(\cdot)$ also satisfies some properties quantifying mixedness:

(i) $M(\rho)=0$  if $\rho$ is pure.

(ii) $M(\rho)=1$  if $\rho=\frac{1}{d}I$ is a maximum mixed state, here $I$ is the identity operator.

(iii) $M(\cdot)$ is concave.

\vspace{0.1in}

\textbf{Theorem 2.} {\it For any  normalized coherence measure $C$, $$C(\rho)+D(\rho)+M(\rho)=1, \ \eqno{(19)}$$
where $C(\rho)$ quantifies the quantum wave feature,
$D(\rho)=1-C(|\rho\rangle\langle\rho|)$ quantifies the quantum particle
feature,  and $M(\rho)=C(|\rho\rangle\langle\rho|)-C(\rho)$ quantifies the mixing of an d-dimensional
quantum state $\rho$, with $M(\rho)=0$ for pure states. }
\vspace{0.1in}

In the conventional approach for quantifying WPD, particle behaviour and wave behaviour
are constrained by the inequality, such as
the elegant work by Greenberger and Yasin \cite{Greenberger}: $D^2+V^2\leq 1.$ Our Theorem 2 introduces a
missing element-mixedness which leads to a more
comprehensive conservation law.
Theorem 2 also suggests that increasing the mixedness of quantum states  will reduce the WPD.
By varying the coherence measures, we can obtain conservation laws from different aspects of coherence.

Let $C$ be a coherence measure from convex roof construction,  the relation (19) becomes
the neat wave-particle-mixedness triality proposed  by Y. Tsui and
S. Kim \cite{Kim}:
$$C_f (\rho) +D_f (\rho) + M_f (\rho)=1, \ \eqno{(20)}$$ where $C_f(\rho)=\min_{p_j,|\psi_j\rangle}\sum_jp_jf(|\psi_j\rangle)$ is the quantification of
the quantum wave feature, $D_f(\rho)=1-C_f(|\rho\rangle\langle\rho|)$ is the quantification of the quantum particle
feature, $M_f (\rho)=C_f(|\rho\rangle\langle\rho|)-C_f(\rho)$ quantifies the mixing of an d-dimensional
quantum state $\rho$, with $M_f(\rho)=0$ for pure states, and
$f$ is some normalized symmetric concave function.

By employing pictorial representations,  we will show  complementary relationship of wave-particle-mixedness triality under the $l_1$-norm and relative entropy of coherence respectively.

Example 1. For a two-path interference with the amplitudes $p$ and $1-p$,  we have $${\mathcal C}_{l_1}(\rho)=2\sqrt{p(1-p)}|\langle\eta_1|\eta_2\rangle| ,\eqno {(21)}$$ $$D(\rho)=1-2\sqrt{p(1-p)}, \eqno {(22)}$$ $$M(\rho)=2\sqrt{p(1-p)}(1-|\langle\eta_1|\eta_2\rangle|).\eqno {(23)}$$
Fig. 1 displays the complementary behaviour of wave feature, particle feature and degree of mixedness of quantum states with $p$, here we set $|\langle\eta_1|\eta_2\rangle|=\frac{1}{3}$.
Specially, if $p=\frac{1}{2}$,
we have $${\mathcal C}_{l_1}(\rho)=|\langle\eta_1|\eta_2\rangle| \eqno {(24)}.$$  However, for a double-slit experiment, the conventional  interference visibility  defined as $${\mathcal V}(\rho)=\frac{I_{Max}(\rho)-I_{Min}(\rho)}{I_{Max}(\rho)+I_{Min}(\rho)}, \eqno {(25)}$$ where $I_{Max}(\rho)$ and $I_{Min}(\rho)$ represent the maximum and minimum intensity \cite{Bimonte, Qureshi}, is just $|\langle\eta_1|\eta_2\rangle|$. Thus fringe visibility
is equal to ${\mathcal C}_{l_1}(\rho)$.

\begin{figure}[htbp]
	\centering
\includegraphics[scale=0.4]{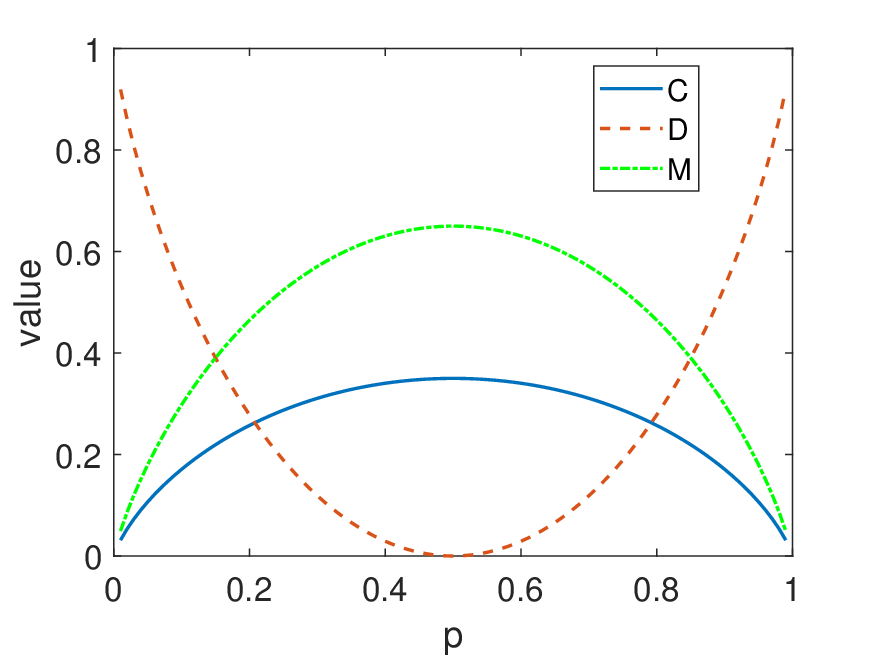}
 \caption{\small The complementary relationship of wave-particle-mixedness triality under ${\mathcal C}_{l_1}$. }
\end{figure}

Example 2. For a two-path interference with the amplitudes $p$ and $1-p$,  we consider $\rho$ in its Bloch representation,
 $\rho=\frac{1}{2}(I+\sum_{i=1}^3 r_i\sigma_i),$ where $r_1=2\sqrt{p(1-p)}\ \text{Re}\langle\eta_2|\eta_1\rangle, r_2=2\sqrt{p(1-p)}\ \text{Im}\langle\eta_2|\eta_1\rangle, r_3=2p-1$,
 $\sigma_1=\left(\begin{array}{cc}
               0 & 1\\
               1 & 0\end{array}\right)$, $\sigma_2=\left(\begin{array}{cc}
               0 & -i\\
               i & 0\end{array}\right)$, $\sigma_3=\left(\begin{array}{cc}
               1 & 0\\
               0 & -1\end{array}\right)$ are Pauli matrices,
               $I$ is the identity operator. Let $r=\sqrt{r_1^2+r_2^2+r_3^2}\leq 1$, the eigenvalues of $\rho$ are $\lambda_1=\frac{1}{2}(1+r)$ and $\lambda_2=\frac{1}{2}(1-r)$, a direct computation shows that
$$\begin{array}{ll}{\mathcal C}_{Re}(\rho)=-p\log p-(1-p)\log {(1-p)}&\\
+\frac{1}{2}(1+r)\log {\frac{1}{2}(1+r)}+\frac{1}{2}(1-r)\log {\frac{1}{2}(1-r)},&\end{array}\eqno {(26)}$$ $$D(\rho)=1+p\log p+(1-p)\log {(1-p)}, \eqno {(27)}$$ $$M(\rho)=1-{\mathcal C}_{Re}(\rho)-D(\rho).\eqno {(28)}$$ Fig. 2 illustrates the complementary behaviour of  wave-particle-mixedness triality of quantum states with $p$, here we  set $\langle\eta_1|\eta_2\rangle=\frac{2}{3}$ and so $r=\sqrt{\frac{16p(1-p)}{9}+(2p-1)^2}.$

\begin{figure}[htbp]
	\centering
\includegraphics[scale=0.4]{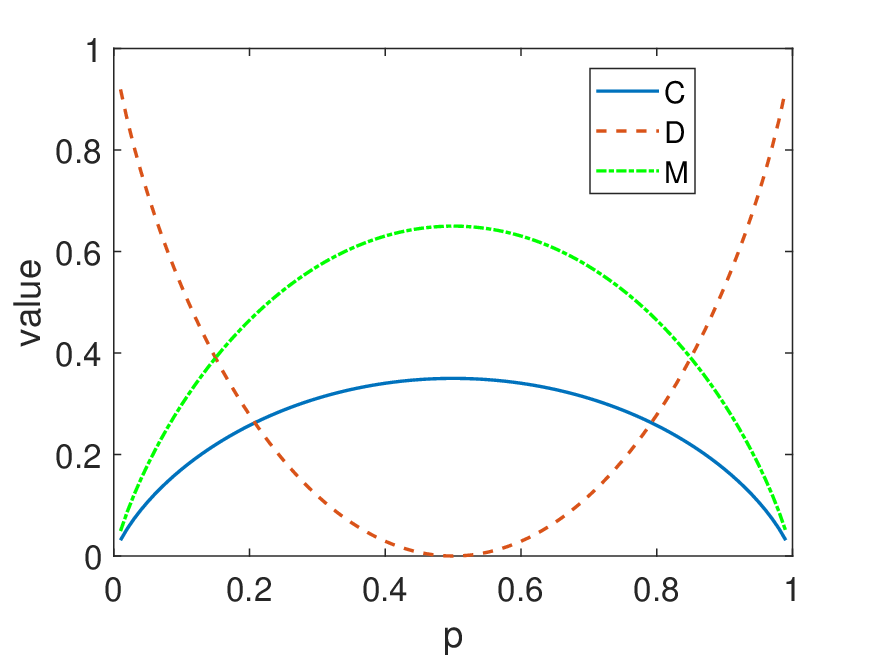}
 \caption{\small The complementary relationship of wave-particle-mixedness triality under $C_{Re}$. }
\end{figure}

%Next, we consider wave-particle-mixedness triality of Werner states under  the $l_1$-norm of coherence.
%Recall that Werner states have the form $$W=\frac{d-m}{d^3-d}I_d\otimes I_d+\frac{dp-1}{d^3-d}F,\ m\in[0,1]$$ where $F=\sum_{1\leq i,j\leq d}|i\rangle\langle j|\otimes |j\rangle\langle i|$ is the flip operator with $\{|i\rangle: \ i=1,2,\ldots, d\}$ an orthonormal basis of ${\mathcal H}$.
%By \cite{Luo3}, $W$ have spectral decomposition $$W=\frac{2p}{d^2+d}II_s+\frac{2(1-p)}{d^2-d}II_t,\ p\in[0,\frac{1}{2}],$$ with eigenvalues
%$\lambda_1=\frac{2p}{d^2+d}$ and $\lambda_2=\frac{2(1-p)}{d^2-d}$ of multiplicities $\frac{d^2+d}{2}$ and $\frac{d^2-d}{2}$, respectively.
%The eigenvalues of
%To be concise, we only treat the case $d=2$. By Theorem 2, a direct computation shows that
%$${\mathcal C}_{l_1}(\rho)=\frac{|2m-1|}{9},$$
%$$D(\rho)=\frac{2}{3}-\frac{4}{9}\sqrt{(m+1)(2-m)},$$
%$$M(\rho)=\frac{1}{3}+\frac{4}{9}\sqrt{(m+1)(2-m)}-\frac{|2m-1|}{9}.\eqno{(23)}$$
%When $m$ is near to  $\frac{1}{2}$, both ${\mathcal C}_{l_1}(\rho)$ and $M(\rho)$ approximate $0$,  $M(\rho)$ approximates $1$,
%this is uniform with maximally mixed property of Werner states.
%Fig.2 shows the behaviour of $C, D$ and $M$ with $m$.

%\begin{figure}[htbp]
	%\centering
%\includegraphics[scale=0.4]{f2.eps}
% \caption{\small The complementary relationship of $2\times 2$ Werner states. }
%\end{figure}
\vspace{0.1in}

$Discussions.$ We build a measure-independent formula of  WPD via coherence in d-path interferometers. Inherently, we find the inner relation between
WPD and quantum coherence and establish a unified  trade-off relation of WPD. This answers affirmatively the question whether there exists a generic
complementary relation which is suitable for all coherence measures \cite{Kim}. Based on this result, we also  provide a measure-independent description of wave-particle-mixedness triality in d-path interferometers.

By our results, any coherence measure (satisfying (C1), (C2b) and (C3)) can induce a complementary relationship.   Therefore, besides the $l_1$-norm, the relative entropy  and convex roof construction of coherence, one can induce complementary relationships by other key coherence measures, such as the geometric coherence \cite{Streltsov}, the coherence cost \cite{Yuan,Winter}, the robustness of coherence \cite{Napoli,Piani}, and the Tsallis $1$-relative entropy of coherence \cite{Rastegin}.
Note that (C3) and (C2b) together imply (C2a),
our results raise one interesting question naturally whether coherence measures satisfying (C1), (C2a) and (C3) can also induce a dual relation of WPD.
This is important for coherence measures induced by the Tsallis $\alpha$-relative entropy \cite{Rastegin} and the Renyi $\alpha$-relative entropy \cite{SLLX}, because these measures do not satisfy (C2b) in general while they fulfill condition (C2a).
Specifically, the Renyi relative $\alpha-$entropies of coherence fulfills (C1), (C2a) and (C3) $(\alpha\in[0,1))$, the Tsallis relative $\alpha-$entropies of coherence fulfills  (C1), (C2a) and (C3) $(\alpha\in(0,2])$.
It is worth studying since it can capture more aspects in exploring the relation between interferometric visibility
and quantum coherence.

We also remark that coherence measures depend  on free operations \cite{Adesso}. The definition of free operations for the resource theory of
coherence is not unique and different choices, often motivated by suitable practical considerations, are being examined in the
literature, such as incoherent operations (IOs) \cite{Baumgratz}, strictly incoherent operations (SIOs) \cite{Winter,Yadin},  physically incoherent operation (PIOs) \cite{Gour}, and genuinely incoherent operations (GIOs) \cite{Vicente}. However, by the hierarchical relationship between IOs, SIOs,
PIOs, and GIOs: $$\text {GIOs}\subseteq \text {PIOs} \subseteq \text {SIOs} \subseteq \text {IOs},\eqno{(29)}$$
and  structural  characterization  of maximally coherent states under genuinely incoherent operations \cite{Du2}, our results  hold true for coherence  measures defined  in above different free  operations.

Our results may be used in the discrete-time quantum walk on the line. Due to limited space, we just give ideas to possible application of our results.
Let us recall the definition of a standard discrete-time quantum walk on the line \cite{Carneiro,AbalR}.
The Hilbert space ${\mathcal H}={\mathcal H}_P\otimes {\mathcal H}_C$ is composed of two parts: a spatial subspace
${\mathcal H}_P$ spanned by the orthonormal set $\{|x\rangle\}$, where the integers $x=0, \pm 1, \pm 2, \cdots$, are associated to discrete positions on the line, and a single-qubit coin space ${\mathcal H}_C$ spanned by two orthonormal vectors denoted by $\{|R\rangle, |L\rangle\}$. A generic state for the walker is $$|\Psi\rangle=\sum_{x=-\infty}^{+\infty} |x\rangle\otimes [a_x|R\rangle+b_x|L\rangle] \eqno{(30)}$$ in terms of complex coefficients satisfying normalization condition $\sum_x(|a_x^2|+|b_x^2|)=1$. A step of the walk is described by the unitary operator
$$U=S(I_P\otimes C), \eqno{(31)}$$ where $C$ is a suitable unitary operation in ${\mathcal H}_C$, $I_P$ is the
identity operator in ${\mathcal H}_P$, the shift operator $$S=S_R\otimes |R\rangle\langle R|+S_L\otimes |L\rangle\langle L| \eqno{(32)}$$
with $S_R=\sum_x |x+1\rangle\langle x|$ and $S_L=S_R^\dag=\sum_x|x-1\rangle\langle x|$. The evolution of an initial state $|\Psi(0)\rangle$  is given by $$|\Psi(t)\rangle=U^t|\Psi(0)\rangle, \eqno{(33)}$$ where the non-negative integer $t$ counts the discrete time steps that have been taken.
We firstly  compute the asymptotic coherence from coherent initial conditions or incoherent initial conditions.
We are aimed to provide an exact expression for the asymptotic (long-time) coherence under special walks, such as the Hadamard walk and so on.
Secondly, we can get the asymptotic (long-time) particle feature during the unitary evolution by Theorem 2.
Since quantum walks are being studied as potential sources for new quantum algorithms, the information of asymptotic coherence and asymptotic particle feature may be helpful for realizing computational advantage arising from coherence \cite{Arnaud,Stahlke}.
A parallel discussion on quantum walks is to characterize the asymptotic entanglement from local or nonlocal initial conditions  \cite{Carneiro,AbalR}.

We also consider the possibility of demonstrating our complementary relationship in experiment and linking our results to one ready made experimental observation. According to our cognition, we need an experimental setup which is engineered to enable the generation of wave-particle transitions, as well as the measurement of properties associated with wave-particle behaviour and mixedness. It is encouraging that wave-particle-entanglement triality has been verified by the silicon-integrated nanophotonic quantum chip \cite{Mafei}. So it may be hopeful to verify our wave-particle-mixedness triad through similar operation.  Specially, correlated photon pairs are generated through three spon-taneous four-wave mixing processes by using a co-polarized bi-chromatic coherent input \cite {Angulo}. We think such correlated photon quantum
states (coherent states) may be prepared as pure quantons entering a d-port interferometer to
perform demonstrations of our complementary relationship in experiment.
In addition, combining our Theorem 2 and the wave-particle-entanglement triality \cite{Mafei}, one can see increasing the mixedness or entanglement of quantum states will reduce the WPD that can be observed. It is harmonic with the observation increased noise leads to a decrease in path predictability \cite{Pathchan} and interference
visibility \cite{Biswas}.

\vspace{0.1in}
{\it Acknowledgement---}
The authors thank referees for providing constructive comments which improve the presentation of the study. The
authors also thank professor Yuanhong Tao for helpful comments
during the preparation of this paper. This research was supported by NSF of Xiamen (3502Z202373018), NSF of China
(12271452), and NSF of Fujian (2023J01028).

\end{document}